\documentclass[11pt]{article}
\usepackage{amssymb,latexsym,amsmath}
\usepackage[dvips]{graphicx}
\usepackage{cite}
\newtheorem{theo}{Theorem}

\newtheorem{lemm}[theo]{Lemma}

\def\nn{\nonumber}

\def\diag{\mathop{\rm diag}\nolimits}
\def\qdots{\mathinner{\mkern1mu\raise1pt\vbox{\kern7pt\hbox{.}}\mkern2mu
 \raise4pt\hbox{.}\mkern2mu\raise7pt\hbox{.}\mkern1mu}}
\def\Z{{\mathbb Z}}

\newcommand{\hatH}{{\hat H}}

\begin{document}
\begin{center}
{\Large \bf
Quantum state transfer in spin chains\\[2mm]
with $q$-deformed interaction terms
}\\[5mm]
{\bf E.I.~Jafarov\footnote{E-mail: ejafarov@physics.ab.az; Permanent address: 
Institute of Physics, Azerbaijan National Academy of Sciences, Javid av.\ 33, AZ-1143 Baku, Azerbaijan} and J.\ Van der Jeugt}\footnote{E-mail:
Joris.VanderJeugt@UGent.be}\\[1mm]
Department of Applied Mathematics and Computer Science,
Ghent University,\\
Krijgslaan 281-S9, B-9000 Gent, Belgium.
\end{center}

\vskip 10mm
\noindent
Short title: spin chains with $q$ deformed interaction


%\addtolength{\baselineskip}{2mm}
%\addtolength{\abovedisplayskip}{1mm}
%\addtolength{\belowdisplayskip}{1mm}
%\addtolength{\parskip}{1mm}

\begin{abstract}
We study the time evolution of a single spin excitation state in certain linear spin chains,
as a model for quantum communication.
Some years ago it was discovered that when the spin chain data (the nearest neighbour interaction strengths and the magnetic field strengths)
are related to the Jacobi matrix entries of Krawtchouk polynomials or dual Hahn polynomials,
so-called perfect state transfer takes place.
The extension of these ideas to other types of discrete orthogonal polynomials did not lead to new models
with perfect state transfer, but did allow more insight in the general computation of the correlation function.
In the present paper, we extend the study to discrete orthogonal polynomials of $q$-hypergeometric type.
A remarkable result is a new analytic model where perfect state transfer is achieved:
this is when the spin chain data are related to the Jacobi matrix of $q$-Krawtchouk polynomials.
The other cases studied here (affine $q$-Krawtchouk polynomials, quantum $q$-Krawtchouk polynomials, dual $q$-Krawtchouk polynomials,
$q$-Hahn polynomials, dual $q$-Hahn polynomials and $q$-Racah polynomials) do not give rise to
models with perfect state transfer. However, the computation of the correlation function itself
is quite interesting, leading to advanced $q$-series manipulations.
\end{abstract}

\section{Introduction}

In quantum information processing, the transfer of a quantum state from one qubit to another qubit is a crucial ingredient.
For long-distance quantum communication, it is generally accepted that this should rely on optical means.
For short-range or mid-range quantum communication (e.g.\ from one quantum processor to another) it seems to be more natural
to use qubit chains (spin chains of interacting fermions) as quantum wires.
In such a chain, qubits interact with their nearest neighbours and the interaction is described by some tridiagonal interaction matrix. 
It is advantageous to accomplish state transfer by just letting the system evolve by itself, without dynamical control.
Bose was the first to propose such spin chains as a quantum channel for quantum state transfer between qubits located
at different ends of a spin chain~\cite{Bose2003,Bose2005}, see~\cite{Bose2007} for a review.
Bose's work inspired many researchers to study the possibility of perfect state transfer in 
spin systems~\cite{Christandl2004,Albanese2004,Christandl2005,Shi2005,PPKF05,Karbach2005,Xi2008,Kay2009}. 

The transmission of quantum states is commonly performed by a chain of qubits coupled via the Heisenberg or
the XY interactions~\cite{DiVicenzo2000,Benjamin2002,Benjamin2003}.
In this context, one often assumes to have individual control of the nearest-neighbour couplings in the spin chain,
leading to the idea of pre-engineered qubit couplings~\cite{Yung2005,Karbach2005}.
An interesting consequence of well-chosen controlled couplings is that one can obtain mirror inversion of a quantum state with
respect to the center of the chain, and this can lead to perfect state transfer~\cite{Christandl2004,Kay2009,Albanese2004}
at certain specified times over arbitrary length of the spin chain. 

Theoretical investigations of spin chains for perfect state transfer can be divided in two classes: analytical solutions and numerical solutions.
The original two systems described in~\cite{Albanese2004} are analytical solutions: the coupling strength $J_k$ at position $k$ of the system ($k=0,1,\ldots,N-1$)
consisting of $N+1$ interacting qubits is some analytic function of $k$ and $N$, and there exist closed form expressions of the 
eigenvalues and eigenstates of the single-fermion states. In the case of~\cite{Albanese2004}, these eigenstates are related to
discrete orthogonal polynomials (namely Krawtchouk polynomials and dual Hahn polynomials).
Another analytical solution was given by Shi {\em et al}~\cite{Shi2005}.
On the other hand, one can use a numerical procedure known as the inverse eigenvalue problem, and design spin chains for 
perfect state transfer numerically~\cite{Karbach2005}. In this case, one starts from a (spatially symmetric) set of values (the
single-fermion energies), and constructs numerically a tridiagonal matrix having these values as eigenvalues. The off-diagonal
elements of this matrix then yield the values for the strengths $J_k$. 

The analytical solutions have a number of advantages, certainly from the mathematical point of view.
Apart from the fact that one has a complete analytic description of the single-fermion eigenvalues and eigenstates (and thus,
through Slater determinant, of all $n$-fermion eigenstates), some other interesting quantities can be computed in closed form.
For example, the transition amplitude at time $t$ of an excitation from site~$s$ to site~$r$ (or the ``correlation function'')
can be determined analytically~\cite{Chakrabarti2010}.

In the two elegant analytical solutions of~\cite{Albanese2004}, the single-fermion interaction matrix is related to the Jacobi
matrix of a system of discrete orthogonal polynomials (Krawtchouk and dual Hahn).
This inspired Chakrabarti and Van der Jeugt~\cite{Chakrabarti2010} to investigate other systems for which the interaction matrix
coincides with the Jacobi matrix of a system of orthogonal polynomials. 
Although no other solutions with perfect state transfer were found this way, the theoretical analysis gave rise to a number of
interesting results: explicit formulae for transition amplitudes (or correlation functions), an explanation of why these two systems
discovered in~\cite{Albanese2004} are so special, and a group theoretical approach of the problem.

Apart from several known systems of discrete orthogonal polynomials of hypergeometric type in the Askey-scheme~\cite{Koekoek},
there is also a list of discrete orthogonal polynomials of $q$-hypergeometric type.
In this context, it is a natural question to ask whether Jacobi matrices of these $q$-orthogonal polynomials could also function
as interaction matrices for spin chains, and whether they would give rise to new solutions with perfect state transfer.
This is the topic treated in the current paper.

Among the main results is indeed a new analytical solution for a spin chain with perfect state transfer.
This new solution occurs in the context of $q$-Krawtchouk polynomials.
We have studied all cases of the $q$-Askey-scheme, and our analysis has shown that this is the only new case with perfect state transfer.
Therefore, we shall concentrate on this $q$-Krawtchouk case, and present all details related to this.
For the other cases the computations and results are rather complicated, and we shall only give a brief summary.

\section{The qubit chain as a spin chain}

In this short section, we shall describe the notation and necessary ingredients for a qubit chain.
This is a system of $N+1$ interacting qubits (spin $1/2$ particles) in a quantum register, with an isotropic Hamiltonian of XY type:
\begin{equation}
\hat H=\frac12 \sum_{k=0}^{N-1} J_k(\sigma^x_k\cdot\sigma^x_{k+1}+\sigma^y_{k+1}
\cdot\sigma^y_{k}) + 
\frac12 \sum_{k=0}^N h_k (\sigma^z_k+1),
\label{Ham1}
\end{equation}
where $J_k$ is the coupling strength between the qubits located at sites $k$ and $k+1$,
and $h_k$ is the ``Zeeman'' energy of a qubit at site~$k$. 
To describe the Hilbert space associated with the Hamiltonian, one adopts a standard
fermionization technique~\cite{Lieb1968}. 
The Jordan-Wigner transformation~\cite{Jordan1928} maps the Pauli matrices to spinless lattice fermions 
$\{a_{k}, a_{k}^{\dagger}|\;k = 0, 1, \ldots, N\}$ obeying the common anticommutation relations,
and in terms of these the Hamiltonian~\eqref{Ham1} takes the form:
\begin{equation}
\hat H= \sum_{k=0}^{N-1} J_k(a_k^\dagger a_{k+1}+a_{k+1}^\dagger a_k) + \sum_{k=0}^N h_k a^\dagger_k a_k .
\label{Ham2}
\end{equation}
This describes a set of $N+1$ fermions on a chain with 
nearest-neighbour interaction (hopping between adjacent sites of the chain), and subject to a non-uniform background
magnetic field denoted by $h_k$ ($k=0,1,\ldots,N$).
Initially, the system is in its completely polarized  
ground state $|{\bf 0}\rangle=|00\cdots0\rangle=|0\rangle \otimes
|0\rangle\otimes\cdots\otimes|0\rangle$, where $|0\rangle$ denotes the spin down state. 
Let $|k)=|00\cdots 010\cdots 0\rangle=a_k^\dagger|{\bf 0}\rangle$ ($k=0,1,\ldots,N$) 
denote a state in which there is a single fermion at the site~$k$
and all other sites are empty, i.e.\ $|k)$ describes the state in which the spin at the
site~$k$ has been flipped to $|1\rangle$. 
Clearly, the set of states $|k)$ ($k=0,1,\ldots,N)$ forms a basis for the single-fermion states of
the system.
In this single-fermion basis, the Hamiltonian $\hatH$ takes the matrix form
\begin{equation}
M=\left(
\begin{array}{ccccc}
h_0 & J_0& 0 & \cdots & 0 \\
J_0 & h_1 & J_1 & \cdots & 0\\
0 & J_1 & h_2 & \ddots &  \\
\vdots & \vdots & \ddots & \ddots& J_{N-1}\\
0 & 0 &  & J_{N-1} & h_N
\end{array}
\right).
\label{Ham-M}
\end{equation}
The dynamics (time evolution) of the system is completely determined by the eigenvalues $\epsilon_j$
and eigenvectors $\varphi_j$ of this matrix. 
It is then a standard technique~\cite{Lieb1968,Albanese2004} 
to describe the $n$-fermion eigenstates of $\hatH$ ($n\leq N$) using the
single-fermion eigenstates $\varphi_j$ and Slater determinants, which is why we concentrate on the single-fermion eigenstates.

The matrix $M$ in~\eqref{Ham-M} is real and symmetric, so the spectral theorem~\cite{Golub1996}
implies that it can be written as $M=UDU^T$,
where $D$ is a diagonal matrix and $U$ an orthogonal matrix:
\begin{align}
& D= \diag (\epsilon_0,\epsilon_1,\ldots, \epsilon_N),\\
& U U^T=U^TU=I.
\end{align}
The entries of $D$ are the single-fermion energy eigenvalues, and the columns of the matrix $U$
are the (orthonormal) eigenvectors of $M$, i.e.\ the single-fermion eigenstates
$\varphi_j= \sum_{k=0}^N U_{kj}\;|k)$ with $\hatH\varphi_j = M\varphi_j = \epsilon_j\,\varphi_j$.

The dynamics of the system under consideration is described by the unitary time evolution operator 
$\mathcal{U}(t) \equiv \exp(-it\hatH)$.
Assume that the ``state sender'' is located at site $s$ of the spin chain, and the ``state receiver''
at site $r$ ($s$ and $r$ are site labels, belonging to $\{0,1,\ldots,N\}$). 
At time $t=0$ the sender turns the system into the single spin state $|s)$.
After a certain time~$t$, the system evolves to the state $\mathcal{U}(t)|s)$ 
which may be expressed as a linear superposition of all the single spin states.
So the transition amplitude of an excitation from site $s$ to site $r$ of the 
spin chain is given by the time-dependent correlation function
\begin{equation}
f_{r,s}(t) = (r|\mathcal{U}(t)|s).
\label{frs}
\end{equation}
Using the orthogonality of the states $\varphi_j$, one finds~\cite{Chakrabarti2010}:
\begin{equation}
f_{r,s}(t) =  \sum_{j=0}^N U_{rj}U_{sj} e^{-it\epsilon_j}, \qquad\hbox{or}\quad
f_{r,s}(t) = \sum_{j=0}^N U_{rj}U_{sj} z^{\epsilon_j} \qquad(z=e^{-it}).
\label{frsz}
\end{equation}
One says that there is {\em perfect state transfer} at time $t$ from one end of the chain to the other end
when $|f_{N,0}(t)|=1$. 
The conditions for perfect state transfer can quite easily be described in terms of the ``mirror symmetry''
of the matrix $M$ in~\eqref{Ham-M}, see~\cite{Karbach2005,Kay2009}.
However, if our aim is to study analytical solutions, we should also require the conditions that
the eigenvalues $\epsilon_j$ and the eigenvector components $U_{kj}$ should be analytic (closed form)
expressions.

We shall now consider the cases where the values
characterizing the system ($J_k$ and $h_k$) are related to the Jacobi matrix
of a set of discrete orthogonal polynomials of $q$-hypergeometric type. 
This has the advantage that the quantities $\epsilon_j$ and $U_{kj}$ are known explicitly,
and from these the correlation functions $f_{r,s}(t)$ can be computed.
Our interest goes beyond perfect state transfer only: the aim is to study those cases
where the correlation function has a closed form expression.
Of course, the cases with perfect state transfer deserve extra attention.

\section{$q$-Krawtchouk polynomials and perfect state transfer}

The purpose of this section is to describe in detail the first case of an interaction matrix of the form~\eqref{Ham-M}
related to the Jacobi matrix of a finite system of discrete orthogonal polynomials of $q$-hypergeometric type.
We shall recall some of the necessary notation for $q$-series.
For the case of $q$-Krawtchouk polynomials, we give the eigenvalues $\epsilon_j$ and eigenvectors $\varphi_j$.
Then the purpose is to compute the correlation function~\eqref{frsz}.
Due to the nonlinearity of $\epsilon_j$ (with respect to $j$), this function turns out to have a complicated structure.
Only for certain forms of the deformation parameter~$q$, and at specific times~$t$, one can simplify the expression of 
the correlation function $f_{r,s}(t)$.
Fortunately, when $q$ is a positive rational number (quotient of two odd integers), and when the other parameter $p$
appearing in the $q$-Krawtchouk polynomials takes the special form $p=q^{-N}$, the system yields perfect state transfer.
This situation is summarized in subsection~3.3.
We end this section by reconsidering the general correlation function, and apply some $q$-series manipulations
in order to write it in an appropriate form. 

\subsection{Notation for $q$-functions and $q$-Krawtchouk polynomials}

The standard reference book on $q$-hypergeometric functions is~\cite{Gasper}, and we follow the notation from this book.
For a list of orthogonal polynomials of $q$-hypergeometric type, see~\cite{Koekoek}.

In the context of $q$-series, $q$ is a positive real number with $q\ne 1$, and for us it can be considered as an extra parameter in the model.
We use the common notation for $q$-numbers:
\begin{equation}
[n] \equiv [n]_q = \frac{1-q^n}{1-q} \qquad (n\in\Z)
\end{equation}
and $[n]\to n$ in the limit $q\to 1$.
For any complex number $a$ and any nonnegative integer $n$, the $q$-shifted factorial is defined by
\begin{equation}
(a;q)_n=(1-a)(1-aq)\cdots(1-aq^{n-1}),
\end{equation}
and the product is just 1 when $n=0$. 
Sometimes, in the context when $0<q<1$, one also uses the infinite product
\begin{equation}
(a;q)_\infty=\prod_{k=0}^\infty (1-aq^k).
\end{equation}
For products of $q$-shifted factorials, it is common to use the abbreviation
\begin{equation}
(a_1,a_2,\ldots,a_A;q)_n = (a_1;q)_n (a_2;q)_n \cdots (a_A;q)_n.
\end{equation}

The $q$-hypergeometric series or {\em basic hypergeometric series} ${}_A\phi_B$ depends on $A$ numerator parameters $a_i$, 
$B$ denominator parameters $b_i$ and a variable $z$ and is defined as~\cite{Gasper}:
\begin{equation}
{}_A\phi_B \left[ \begin{array}{c} a_1, a_2,\ldots,a_A \\ b_1,\ldots, b_B \end{array} ; q,z\right] =
\sum_{n=0}^\infty \frac{ (a_1,a_2,\ldots,a_A;q)_n }{(q,b_1,\ldots,b_B;q)_n} \left[ (-1)^n q^{\binom{n}{2}}\right]^{1+B-A} z^n.
\label{q-hyp}
\end{equation}
In most of the series considered here, the numerator contains a parameter of the form $q^{-m}$, with $m$ a nonnegative integer.
In that case, the series~\eqref{q-hyp} is {\em terminating}: it has only $m+1$ terms since $(q^{-m};q)_n=0$ for $n=m+1,m+2,\ldots$.
If one of the denominator parameters is of the form $q^{-N}$ (with $N$ a positive integer), then one of the numerator parameters
should be of the form $q^{-m}$ with $m\leq N$ in order to make sure that the series terminates before one reaches zeros in the denominator.
This will always be the case here.

Let us now consider $q$-Krawtchouk polynomials $K_n \left( {q^{ - x} ;p,N;q} \right)$, 
characterized by a positive integer parameter $N$ and a positive real parameter $p$: $p>0$. 
This polynomial of degree $n$ in $q^{-x}$ is defined as~\cite{Koekoek}
\begin{equation}
K_n \left( {q^{ - x} } \right) \equiv K_n({q^{ - x} ;p,N;q}) = 
\,{}_3\phi_2 \left[ \begin{array}{c} q^{-n} ,q^{-x} , -pq^n \\ q^{-N} ,0 \end{array}; q,q \right], \qquad n = 0,1, \ldots ,N.
\end{equation}
The $q$-Krawtchouk polynomials satisfy a discrete orthogonality relation, namely
\begin{equation}
\sum_{x =0}^N w(x) K_m(q^{-x}) K_n(q^{-x})  = d_n \delta _{mn} ,
\label{orth-K}
\end{equation}
where the weight function is
\begin{equation}
w(x) = \frac{(q^{-N} ;q)_x}{(q;q)_x} (-p)^{-x} ,
\end{equation}
and the square norm takes the rather complicated form
\begin{equation}
d_n  = \frac{(q, -pq^{N+1};q)_n}{(-p,q^{-N};q)_n} \frac{(1+p)}{(1+pq^{2n})} (-pq;q)_N \,p^{-N} q^{-\binom{N+1}{2}} (-pq^{-N})^n q^{n^2} .
\label{dn}
\end{equation}
It is easy to see that $d_n>0$ for $0<q<1$ and also for $q>1$ (since $p>0$).
The polynomials $K_n(q^{-x})$ also satisfy the following three term recurrence relation:
\begin{equation}
 -(1-q^{-x})K_n(q^{-x}) = A_n K_{n+1}(q^{-x}) - (A_n+C_n)K_n(q^{-x}) + A_n K_{n-1}(q^{-x}),
\end{equation}
with~\cite{Koekoek}
\begin{align*}
A_n  &= \frac{(1-q^{n-N})(1+pq^n)}{(1+pq^{2n})(1+pq^{2n+1})},\\
C_n  &=  - pq^{2n-N-1} \frac{(1+pq^{n+N})(1-q^n)}{(1+pq^{2n-1})(1+pq^{2n})}.
\end{align*}
It is appropriate to introduce orthonormal $q$-Krawtchouk functions
\begin{equation}
\tilde K_n(q^{-x}) \equiv \sqrt{\frac{w(x)}{d_n}} K_n(q^{-x})
\end{equation}
and then the corresponding orthonormal recurrence relation is
\begin{equation}
-[-x]\tilde K_n(q^{-x})=- J_{n-1} \tilde K_{n-1}(q^{-x})+ h_n \tilde K_n(q^{-x})-J_n \tilde K_{n+1}(q^{-x}) ,
\end{equation}
where
\begin{equation}
J_n = -\frac{A_n}{1-q}\sqrt{\frac{d_{n+1}}{d_n}} ,\qquad h_n = -\frac{A_n+C_n}{1-q} .
\label{qK-J-h}
\end{equation}
Again, it is easy to verify that $J_n>0$ ($n=0,1,\ldots,N-1$) and $h_n>0$ ($n=0,1,\ldots,N$).
Following the technique of~\cite{Chakrabarti2010} and~\cite{Regniers2009}, we now have
\begin{lemm}
\label{lemm1}
Let $M_{qK}$ be the tridiagonal $(N+1)\times(N+1)$-matrix (Jacobi matrix)
\begin{equation}
\label{MK}
M_{qK}= \left( \begin{array}{ccccc}
             h_0 & -J_0  &    0   &        &      \\
            -J_0 &  h_1  &  -J_1  & \ddots &      \\
              0  & -J_1  &   h_2  & \ddots &  0   \\
                 &\ddots & \ddots & \ddots & -J_{N-1} \\
                 &       &    0   &  -J_{N-1}  &  h_N
          \end{array} \right)
\end{equation}
where $J_n$ and $h_n$ are given by~\eqref{qK-J-h},
and let $U$ be the $(N+1)\times(N+1)$-matrix with elements $U_{jk}=\tilde K_j(q^{-k})$.
Then 
\begin{equation}
U U^T = U^TU=I \qquad\hbox{and}\qquad M_{qK}=UDU^T
\end{equation}
where
\begin{equation}
D= \diag (\epsilon_0,\epsilon_1,\epsilon_2, \ldots,\epsilon_N).
\end{equation}
Herein,
\begin{equation}
\epsilon_j = -[-j] = -\frac{1-q^{-j}}{1-q}=q^{-1}+q^{-2}+\cdots+q^{-j}.
\label{eigvals}
\end{equation}
\end{lemm}
In other words, the eigenvectors of the Hamiltonian (in the single-fermion case) corresponding
to the quantities~\eqref{qK-J-h} have components
equal to normalized $q$-Krawtchouk polynomials, and the corresponding energy eigenvalues are $\epsilon_j$
($j=0,1,\ldots,N$). 

Finally, note that working with all positive $J_n$-values as in~\eqref{Ham-M}, or with a matrix like~\eqref{MK} where all
off-diagonal elements are negative, does not make an essential difference (see the remark at the end of section~1 in~\cite{Chakrabarti2010}).
The main difference is that there are also sign changes in the components of the corresponding eigenvectors.

\subsection{Computation of the correlation function}

Consider now a spin chain for which the values of $J_n$ and $h_n$ are fixed by~\eqref{qK-J-h}.
For this system, the eigenvalues of the single-fermion states are given by~\eqref{eigvals},
and the eigenvectors by $\varphi_j= \sum_{k=0}^N U_{kj}\;|k)$ with $U_{jk}=\tilde K_j(q^{-k})$.

In this spin chain, consider the transition from a sender site $s$ to a receiver site $r$.
The transition amplitude or correlation function, given in general by~\eqref{frsz}, becomes:
\begin{align}
f_{r,s}(t) &= \sum_{k=0}^N U_{rk}U_{sk}z^{\epsilon_k} = \sum_{k=0}^N \tilde K_r(q^{-k}) \tilde K_s(q^{-k}) z^{-[-k]} \nn\\
 &= \frac{1}{\sqrt{d_r d_s}} \sum_{k=0}^N w(k) K_r(q^{-k}) K_s(q^{-k}) z^{-[-k]} \qquad (z=e^{-it}).
\label{sumK}
\end{align}
The purpose is to investigate whether one can compute the sum in~\eqref{sumK}. 
For this, let us first concentrate on the special case when sender and receiver are at different ends of the chain, namely when $s=0$ and $r=N$.
After some simplifications, one finds for the summation part:
\begin{equation}
\sum_{k=0}^N w(k) K_N(q^{-k}) K_0(q^{-k}) z^{-[-k]} = 
\sum_{k=0}^N \frac{(q^{-N};q)_k}{(q;q)_k} q^{Nk} z^{-[-k]}.
\label{sumpart}
\end{equation}
Because of the factor $z^{-[-k]}$, the sum in~\eqref{sumpart} is not of $q$-hypergeometric type, hence there is no hope that it can
be simplified any further for arbitrary values of~$z$ (i.e.\ for arbitrary values of $t$).
For certain specific values, however, simplification does take place.
Since $z=e^{-it}$, one has
\[
z^{-[-k]} = e^{-it(q^{-1}+q^{-2}+\cdots+q^{-k})}.
\]
Assume now that the deformation parameter $q$ is a rational number of the following form:
\begin{equation}
q^{-1}=\frac{P}{Q}, \quad\hbox{with $P$ and $Q$ {\em odd} positive integers (having no common factors)}.
\end{equation}
Then, for each value of the index $k$ with $k\leq N$:
\begin{align}
q^{-1}+q^{-2}+\cdots+q^{-k} &= \frac{1}{Q^N}(PQ^{N-1}+P^2Q^{N-2}+\cdots+P^kQ^{N-k}) \nn\\
&= \frac{1}{Q^N} \times k\times\hbox{(an odd integer)}.
\end{align}
Suppose now that we consider the system at time 
\[
t=T\equiv Q^N\pi,
\]
then
\[
z^{-[-k]} = e^{-it(q^{-1}+q^{-2}+\cdots+q^{-k})}=e^{-i\pi(PQ^{N-1}+P^2Q^{N-2}+\cdots+P^kQ^{N-k})}
= (-1)^k.
\]
In this case, the expression~\eqref{sumpart} simplifies drastically, since
\begin{equation}
\sum_{k=0}^N \frac{(q^{-N};q)_k}{(q;q)_k} q^{Nk} (-1)^k= (-1;q)_N = \prod_{j=0}^{N-1} (1+q^j).
\label{sum-1}
\end{equation}
This is in fact a consequence of the $q$-binomial theorem~\cite[(II.4)]{Gasper}:
\begin{equation}
\sum_{k=0}^N \frac{(q^{-N};q)_k}{(q;q)_k} q^{Nk} x^k= (x;q)_N.
\label{sum-2}
\end{equation}
Taking into account the expressions for $d_0$ and $d_N$ from~\eqref{dn}, 
one has
\[
f_{N,0}(T) = \frac{(-1;q)_N}{\sqrt{d_0 d_N}} = (-1;q)_N \sqrt{ \frac{p^N q^{N(N+1)/2}}{(-pq,-pq^N;q)_N} }
\]
It is not too difficult to see that this expression takes its maximum value when $p=q^{-N}$, and in that case
\[
f_{N,0}(T)\Big\vert_{p=q^{-N}} = \sqrt{ \frac{q^{-N(N-1)/2}(-1;q)_N}{(-q^{-N+1};q)_N }} = 1,
\]
and thus we have discovered a new analytic model with perfect state transfer.

Note that at time $t=2T$ one has $z^{-[-k]}=1$ for all $k$, and thus from~\eqref{sum-2} we have $f_{N,0}(2T)=0$.
More generally, in that case it simply follows from the orthogonality of $q$-Krawtchouk polynomials and~\eqref{sumK}
that $f_{r,s}(2T)=\delta_{r,s}$. Clearly, the system is periodic in time with period $2T$.

\subsection{Perfect state transfer related to $q$-Krawtchouk polynomials}

It is convenient to collect all the ingredients for the new model in this subparagraph.
Perfect state transfer takes place if the parameter $p$ appearing in $q$-Krawtchouk polynomials
is equal to $q^{-N}$: $p=q^{-N}$. In that case, $w(x)$ and $d_n$ take the simpler form
\[
w(x)=\frac{(q;q)_N}{(q;q)_x(q;q)_{N-x}} q^{x(x-1)/2}
\]
and
\[
d_n= 2 \frac{(q,-q;q)_n(q,-q;q)_{N-n}}{(q;q)_N (q^n+q^{N-n})}.
\]
After some straightforward simplifications, one finds for the spin chain data $J_n$ ($n=0,1,\ldots,N-1$) and
$h_n$ ($n=0,1,\ldots,N$) from~\eqref{qK-J-h}:
\begin{equation}
J_n= \sqrt{[n+1][N-n]} \frac{q}{q^{N-n}+q^{n+1}} \left( \frac{(1+q^{N-n})(1+q^{n+1})}{(q^{N-n}+q^{n+2})(q^{N-n+1}+q^{n+1})}\right)^{1/2},
\label{Jn}
\end{equation}
and
\begin{equation}
h_n= [n] \frac{(1+q^n)}{(q^{N-n}+q^n)(q^{N-n+1}+q^n)} + [N-n] \frac{(1+q^{N-n})}{(q^{N-n}+q^n)(q^{N-n}+q^{n+1})}.
\label{hn}
\end{equation}
Note that $h_n=h_{N-n}$ ($n=0,1,\ldots,N$) and $J_{n}=J_{N-1-n}$ ($n=0,1,\ldots,N-1$), so the matrix~\eqref{MK}
is mirror symmetric, as required for perfect state transfer.

We can now state the following result.
\begin{theo}
Let $q\ne 1$ be a positive number of the form $q=Q/P$ with $Q$ and $P$ positive odd integers.
For the spin chain system~\eqref{Ham2} characterized by the couplings~\eqref{Jn} and the $h_n$'s~\eqref{hn}, 
there is perfect state transfer from site~$0$ to site~$N$ at time $t=T=Q^N\pi$. 
At time $2T$ the system is back to its original state (at time~0); in fact the system is periodic in time with period $2T$.
The single fermion eigenvalues are
\[
\epsilon_j = q^{-1}+q^{-2}+\ldots+q^{-j}, \qquad (j=0,1,\ldots,N),
\]
and the single fermion eigenstates are given by $\varphi_j= \sum_{n=0}^N U_{nj}\;|n)$ where
\[
U_{nj}= \sqrt{\frac{w(j)}{d_n}} K_n(q^{-j})= \sqrt{\frac{w(j)}{d_n}}
\,{}_3\phi_2 \left[ \begin{array}{c} q^{-n} ,q^{-j} , -q^{n-N} \\ q^{-N} ,0 \end{array}; q,q \right].
\]
\end{theo}

\subsection{General remarks}

The special case where perfect state transfer is possible occurs when $q^{-1}=P/Q$, with $P$ and $Q$ odd (positive) integers.
What happens in other cases? Clearly, when $P$ and $Q$ would be both even integers, they have a common factor that can be canceled.
So there remains to see what happens when $q^{-1}$ is a rational number of the form
\[
\frac{\hbox{even}}{\hbox{odd}} \qquad \hbox{or} \qquad \frac{\hbox{odd}}{\hbox{even}}.
\]
In that case, one can write:
\begin{equation}
q^{-1} = 2^r\frac{P}{Q} \qquad \hbox{or} \qquad q^{-1} = 2^{-r}\frac{P}{Q},
\label{other}
\end{equation}
where $r$ is a positive integer and $P$ and $Q$ are again odd (positive) integers.
In the first case, we have
\begin{align}
q^{-1}+q^{-2}+\cdots+q^{-k} &= \frac{2^r}{Q^N}(PQ^{N-1}+2^{r}P^2Q^{N-2}+\cdots+2^{r(k-1)}P^kQ^{N-k}) \nn\\
&= \frac{2^r}{Q^N} \times\hbox{(an odd integer)},
\end{align}
so the parity of the expression in brackets becomes independent of~$k$.
In the second case, one has
\begin{align}
q^{-1}+q^{-2}+\cdots+q^{-k} &= \frac{1}{2^{rk}Q^N}(2^{r(k-1)}PQ^{N-1}+\cdots+2^{r}P^{k-1}Q^{N-k+1}+P^kQ^{N-k}) \nn\\
&= \frac{1}{2^{rk}Q^N} \times\hbox{(an odd integer)},
\end{align}
so again the parity of the expression in brackets becomes independent of~$k$.
As a consequence, one can never obtain that $z^{-[k]}=(-1)^k$ (or $z^{-[k]}$ of the form $x^k$) 
for a certain time $t$ in the cases~\eqref{other},
and this is a condition for further simplification of the sum~\eqref{sum-2}. 

Now that is has become clear that $q=Q/P$, with $Q$ and $P$ odd integers, plays a special role, let us go back to 
the general case (i.e.\ arbitrary $p$) with a parameter $q$ of this form.
The expression of the general correlation function is given by~\eqref{sumK}:
\begin{equation}
f_{r,s}(t)=\frac{1}{\sqrt{d_rd_s}} \sum_{k=0}^N \frac{(q^{-N};q)_k (-p)^{-k}}{(q;q)_k} \,
\,{}_3\phi_2 \left[ \begin{array}{c} q^{-k} ,q^{-r} , -pq^r \\ q^{-N} ,0 \end{array}; q,q \right]
\,{}_3\phi_2 \left[ \begin{array}{c} q^{-k} ,q^{-s} , -pq^s \\ q^{-N} ,0 \end{array}; q,q \right]
z^{-[-k]}.
\label{tmp}
\end{equation}
Just as in~\eqref{sum-1}, this simplifies at time $t=T=Q^N\pi$, since then $z^{-[-k]}=(-1)^k$.
The previous summation can then be simplified using the (symmetric) Poisson kernel for Al-Salam--Chihara
polynomials (see~\cite[Proposition~4]{Jagannathan} or~\cite[(14.8)]{ARS}):
\begin{align}
&\sum_{k=0}^\infty \,{}_3\phi_2 \left[ \begin{array}{c} q^{-k}, a,b \\ f ,0 \end{array}; q,q \right]
\,{}_3\phi_2 \left[ \begin{array}{c} q^{-k} ,c,d \\ f ,0 \end{array}; q,q \right]
\frac{(f;q)_k}{(q;q)_k} x^n = \nn\\
&\frac{(abcx,abdx,acdx,bcdx,fx;q)_\infty}{(acx,bcx,adx,bdx,abcdx;q)_\infty}
{\;}_8W_7(abcdxq^{-1};a,b,c,d,abcdx/f;q,fx).
\label{Ser2}
\end{align}
Herein, ${}_8W_7$ is the notation for a very-well-poised ${}_8\phi_7$ series~\cite[\S~2.1]{Gasper}.
In the current case of~\eqref{tmp}, what results is a terminating ${}_8\phi_7$ series
to which Watson's transformation formula~\cite[(III.18)]{Gasper} can be applied. 
This transformation formula rewrites the special ${}_8\phi_7$ series as a terminating ${}_4\phi_3$ series,
and after some simplifications one finds:
\begin{align}
f_{r,s}(T)& =\frac{1}{\sqrt{d_rd_s}}  \frac{(-q^{-s};q)_r (-q^{-r};q)_s (p^{-1}q^{-N};q)_{N-r-s} (q^{-N};q)_{r+s}}{(q^{-N};q)_r (q^{-N};q)_s} \nn\\
& \times\ \,{}_4\phi_3 \left[ \begin{array}{c} q^{-r} ,q^{-s} , pq^N, p^{-1}q^{-r-s} \\ -q^{-r}, -q^{-s}, q^{1+N-r-s} \end{array}; q,q \right].
\label{4phi3}
\end{align}
This is a terminating balanced ${}_4\phi_3$ series, and it can be written in several forms according to Sears' transformation~\cite[(III.15)]{Gasper}.
(The form given here is the most symmetric one, but if $1+N-r-s$ is a negative integer smaller than $-r$ or $-s$, it should be transformed so that
the termination of the series is not caused by the denominator.)
Expression~\eqref{4phi3} is a simple formula for the correlation function with arbitrary $r$ and $s$, arbitrary parameter $p$, but valid only
at time $t=T$ (half the period).

Let us now consider again the case that $p=q^{-N}$. Then~\eqref{4phi3} yields
\begin{equation}
f_{r,s}(T) =\frac{1}{\sqrt{d_rd_s}}  \frac{(-q^{-s};q)_r (-q^{-r};q)_s (1;q)_{N-r-s} (q^{-N};q)_{r+s}}{(q^{-N};q)_r (q^{-N};q)_s}.
\label{frsT}
\end{equation}
So, due to the appearance of $(1;q)_{N-r-s}$ one finds
\begin{equation}
f_{r,s}(T)= \delta_{r+s,N}.
\end{equation}
In other words, in the case $q=Q/P$ and $p=q^{-N}$, there is also perfect state transfer from site~$s$ to site $N-s$ at time $t=T=Q^N\pi$.

\section{Spin chains related to other orthogonal polynomials of $q$-hyper\-geo\-met\-ric type}

The $q$-Krawtchouk polynomials described in the previous section are not the only $q$-defor\-ma\-tions of ordinary Krawtchouk polynomials.
In fact, there are four possible $q$-generalizations, see~\cite{Koekoek}.
The remaining three cases are known as the affine $q$-Krawtchouk polynomials, the quantum $q$-Krawtchouk polynomials and the dual $q$-Krawtchouk polynomials.
All these polynomials satisfy a discrete orthogonality relation with finite support $\{0,1,\ldots,N\}$.
Apart from these $q$-generalizations of Krawtchouk polynomials, there are three other sets of orthogonal polynomials of $q$-hypergeometric type
with a discrete orthogonality over a finite support, namely $q$-Hahn polynomials, dual $q$-Hahn polynomials and $q$-Racah polynomials.
In the context of the present paper, all these cases should be investigated following the techniques and ideas of section~3.
The purpose of the current section is to give an overview of the main results for all these cases.
First of all, the central (but negative) conclusion is that none of these new cases gives rise to an analytic model with perfect state transfer.
In this sense, the example of section~3 is unique.
In view of this, we think it is appropriate to describe not all the details of the computations involved for these remaining six cases.
Although some of the computations of correlation functions lead to interesting $q$-series manipulations, especially for mathematicians
fascinated by $q$-series, we have chosen not to overload this section with such details.
The reader interested in the actual $q$-series computations should find sufficient information in the summary given in this section.
Others can just skip to the final expression of the transition amplitude $f_{N,0}$ for transfer from site $0$ to site $N$ for each
of the following cases, and establish that perfect state transfer is not possible.

\subsection{Affine $q$-Krawtchouk polynomials}

Let $N$ be a fixed positive integer, and $q$ a positive real parameter with $q\ne 1$.
The affine $q$-Krawtchouk polynomials $K^{\hbox{\scriptsize Aff}}(q^{-x};p,N;q)$ are characterized
by a positive real parameter $p$, and defined by~\cite{Koekoek}
\begin{equation}
K^{\hbox{\scriptsize Aff}}_n(q^{-x}) \equiv K^{\hbox{\scriptsize Aff}}_n(q^{-x};p,N;q) = 
\,{}_3\phi_2 \left[ \begin{array}{c} q^{-n} ,q^{-x} , 0 \\ pq, q^{-N} \end{array}; q,q \right], \qquad n = 0,1, \ldots ,N.
\end{equation}
The affine $q$-Krawtchouk polynomials satisfy a discrete orthogonality relation, namely
\begin{equation}
\sum_{x =0}^N w(x) K^{\hbox{\scriptsize Aff}}_m(q^{-x}) K^{\hbox{\scriptsize Aff}}_n(q^{-x})  = d_n \delta _{mn} ,
\label{orth-aff-K}
\end{equation}
where~\cite{Koekoek}
\begin{equation}
w(x) = \frac{(pq;q)_x(q;q)_N}{(q;q)_x(q;q)_{N-x}} (pq)^{-x} , \qquad
d_n  = \frac{(q;q)_n(q;q)_{N-n}}{(pq;q)_n(q;q)_N} (pq)^{n-N} .
\label{aff-dn}
\end{equation}
For positivity, one needs 
\begin{equation}
0<p<q^{-1}\hbox{ when }0<q<1 \qquad\hbox{and}\qquad 0<p<q^{-N}\hbox{ when }q>1, 
\label{cond-pq}
\end{equation}
and we shall assume
that $p$ satisfies this condition.
The orthonormal affine $q$-Krawtchouk functions
\begin{equation}
\tilde K^{\hbox{\scriptsize Aff}}_n(q^{-x}) \equiv \sqrt{\frac{w(x)}{d_n}} K^{\hbox{\scriptsize Aff}}_n(q^{-x})
\end{equation}
satisfy the following recurrence relation, which follows from~\cite[(3.16.3)]{Koekoek}
\begin{equation}
-[-x]\tilde K^{\hbox{\scriptsize Aff}}_n(q^{-x})=
- J_{n-1} \tilde K^{\hbox{\scriptsize Aff}}_{n-1}(q^{-x})+ h_n \tilde K^{\hbox{\scriptsize Aff}}_n(q^{-x})-J_n \tilde K^{\hbox{\scriptsize Aff}}_{n+1}(q^{-x}) ,
\end{equation}
with
\begin{equation}
J_n = -[n-N](1-pq^{n+1})\sqrt{\frac{d_{n+1}}{d_n}} ,\qquad h_n = [n]pq^{n-N}-[n-N](1-pq^{n+1}) ,
\label{aff-qK-J-h}
\end{equation}
where $J_n>0$ ($n=0,1,\ldots,N-1$) and $h_n>0$ ($n=0,1,\ldots,N$).
The same result as Lemma~\ref{lemm1} now applies, but with the spin chain data given by~\eqref{aff-qK-J-h}.

Let us now consider the computation of the transition amplitude in a spin chain governed by the
quantities~\eqref{aff-qK-J-h} $(z=e^{-it})$:
\begin{align}
f_{r,s}(t) &= \frac{1}{\sqrt{d_r d_s}} \sum_{k=0}^N w(k) K^{\hbox{\scriptsize Aff}}_r(q^{-k}) K^{\hbox{\scriptsize Aff}}_s(q^{-k}) z^{-[-k]} \nn\\
&= \frac{1}{\sqrt{d_r d_s}} \sum_{k=0}^N \frac{(pq;q)_k(q;q)_N (pq)^{-k}}{(q;q)_k(q;q)_{N-k}} 
\,{}_3\phi_2 \left[ \begin{array}{c} q^{-k} ,q^{-r} , 0 \\ pq, q^{-N} \end{array}; q,q \right]
\,{}_3\phi_2 \left[ \begin{array}{c} q^{-k} ,q^{-s} , 0 \\ pq, q^{-N} ,0 \end{array}; q,q \right]
z^{-[-k]}.
\label{sum-aff-qK}
\end{align}
As in subsection~3.2, this sum cannot be simplified unless $q^{-1}$ is of the special form $q^{-1}=P/Q$ with $P$ and $Q$ positive odd integers,
and $t=T=Q^N\pi$. For this time $t$, we have $z^{-[-k]}=(-1)^k$, and now one can continue with the computation of~\eqref{sum-aff-qK}.
In this case, the product of the two ${}_3\phi_2$ functions is computed by applying the product formula of $q$-Hahn polynomials
given by~\cite[(8.3.3)]{Gasper} and substituting $a=p$ and $b=0$ in this expression. 
Following this, the product of the two ${}_3\phi_2$ series in~\eqref{sum-aff-qK} can be rewritten as a sum ($\sum_{m=0}^k$) over
terminating ${}_3\phi_3$ series (terminating because $q^{-m}$ appears as a numerator parameter in the ${}_3\phi_3$).
Exchanging summations over $k$ and over $m$, and performing the inner summation over $k$ (which can be simplified using 
the $q$-binomial theorem) finally leads to the following:
\begin{equation}
f_{r,s}(T)= \frac{(-1;q)_N}{\sqrt{d_r d_s }}\sum_{m = 0}^N \frac{(q^{r-N},q^{s-N};q)_m (pq^{r+s})^{-m}}{(q,-q^{1-N},q^{-N};q)_m} 
\, {}_3 \phi_3 \left[ \begin{array}{c} q^{-r},q^{-s},q^{-m} \\ pq,q^{N-r+1-m},q^{N-s+1-m} \end{array}; q,pq^{2N-m+3} \right].
\end{equation}
The current expression can in general not be simplified further, except in the special cases when the sender is at one end
of the chain ($s=0$). One finds:
\begin{equation}
f_{r,0}(T)= \frac{(-1;q)_N}{\sqrt{d_r d_0 }} \, {}_2 \phi_1 \left[ \begin{array}{c} q^{r-N},0 \\ -q^{1-N} \end{array}; q,\frac{1}{pq^r} \right],
\end{equation}
and in particular
\begin{equation}
f_{N,0}(T)= \frac{(-1;q)_N}{\sqrt{d_N d_0 }} = (-1;q)_N\; (pq)^{N/2} \sqrt{(pq;q)_N}.
\end{equation}
It can be verified that this expression never reaches the value 1 (for $N>1$), due to the conditions~\eqref{cond-pq} for $p$.
So perfect state transfer is not possible in this case.
For certain values of $p$ and $q$, so called high-fidelity transfer can still be achieved~\cite{Bose2003}, and in this context 
such models might be worth considering.

\subsection{Quantum $q$-Krawtchouk polynomials}

The quantum $q$-Krawtchouk polynomials $K^{\hbox{\scriptsize qtm}}(q^{-x};p,N;q)$ are characterized
by a positive real parameter $p$, and defined by~\cite{Koekoek}
\begin{equation}
K^{\hbox{\scriptsize qtm}}_n(q^{-x}) \equiv K^{\hbox{\scriptsize qtm}}_n(q^{-x};p,N;q) = 
\,{}_2\phi_1 \left[ \begin{array}{c} q^{-n} ,q^{-x} \\ q^{-N} \end{array}; q,pq^{n+1} \right], \qquad n = 0,1, \ldots ,N.
\end{equation}
The orthogonality relation is of the form~\eqref{orth-aff-K}, but now with
\begin{align}
w(x) &= \frac{(pq;q)_{N-x}}{(q;q)_x(q;q)_{N-x}} (-1)^{x} q^{x(x-1)/2} , \nn\\
d_n  &= \frac{(q;q)_{N-n}(q,pq;q)_{n}}{(q,q;q)_N} (-1)^{N-n} p^N q^{Nn+N(N+1)/2-n(n+1)/2} .
\label{qtm-dn}
\end{align}
For positivity, one needs 
\begin{equation}
p>q^{-N}\hbox{ when }0<q<1 \qquad\hbox{and}\qquad p>q^{-1}\hbox{ when }q>1.
\label{cond-qtm}
\end{equation}
The orthonormal quantum $q$-Krawtchouk functions
satisfy the following recurrence relation, which follows from~\cite[(3.14.3)]{Koekoek}
\begin{equation}
-[-x]\tilde K^{\hbox{\scriptsize qtm}}_n(q^{-x})=
- J_{n-1} \tilde K^{\hbox{\scriptsize qtm}}_{n-1}(q^{-x})+ h_n \tilde K^{\hbox{\scriptsize qtm}}_n(q^{-x})-J_n \tilde K^{\hbox{\scriptsize qtm}}_{n+1}(q^{-x}) ,
\end{equation}
where
\begin{equation}
J_n = -\frac{[n-N]}{p q^{2n+1}}\sqrt{\frac{d_{n+1}}{d_n}} ,\qquad h_n = -\frac{[n](1-pq^n)}{pq^{2n}}-\frac{[n-N]}{pq^{2n+1}} .
\label{qtm-qK-J-h}
\end{equation}
The computation of the transition amplitude in a spin chain governed by~\eqref{qtm-qK-J-h} is again a rather
technical question. As in the previous subsection, the summation part cannot be simplified unless $z^{-[-k]}=(-1)^k$,
and this happens when $q^{-1}=P/Q$ with $P$ and $Q$ positive odd integers, and $t=T=Q^N\pi$. 
Using similar techniques as in subsection~4.1, one finds:
\begin{align}
f_{r,s}(T)&= (-1)^N \frac{(-1;q)_N}{\sqrt{d_r d_s }}\frac{(pq;q)_N}{(q;q)_N}
\sum_{m=0}^N \frac{(q^{-r},q^{-s};q)_m (pq^{r+s+1-N})^{m}}{(q,-q^{1-N},q^{-N};q)_m} \nn\\
&\times \, {}_3 \phi_3 \left[ \begin{array}{c} q^{r-N},q^{s-N},q^{-m} \\ p^{-1}q^{-N},q^{r-m+1},q^{s-m+1} \end{array}; q,p^{-1}q^{N-m+2} \right].
\end{align}
The current expression can in general not be simplified further, but one has
\begin{equation}
f_{N,0}(T)= (-1;q)_N\; p^{-N} q^{-(3N^2+N)/4} \sqrt{(-1)^N(pq;q)_N}.
\end{equation}
Due to the conditions~\eqref{cond-qtm} this expression never assumes the value 1 (for $N>1$), 
so perfect state transfer is not possible.

\subsection{Dual $q$-Krawtchouk polynomials}

The dual $q$-Krawtchouk polynomials $K(\lambda(x);c,N;q)$ are polynomials of degree~$n$ in
$\lambda(x)=q^{-x}+cq^{x-N}$, characterized by a real parameter $c$ with $c<0$, and defined by~\cite{Koekoek}
\begin{equation}
K_n(\lambda(x)) \equiv K_n(\lambda(x);c,N;q) = 
\,{}_3\phi_2 \left[ \begin{array}{c} q^{-n} ,q^{-x},cq^{x-N} \\ q^{-N},0 \end{array}; q,q \right], \qquad n = 0,1, \ldots ,N.
\end{equation} 
The orthogonality relation is of the form~\eqref{orth-aff-K}, but now with
\begin{equation}
w(x) = \frac{(cq^{-N},q^{-N};q)_x}{(q,cq;q)_x} \frac{1-cq^{2x-N}}{1-cq^{-N}} c^{-x} q^{x(2N-x)} , \qquad
d_n  = \frac{(q;q)_{n}(c^{-1};q)_N}{(q^{-N};q)_n} (cq^{-N})^n .
\label{dual-dn}
\end{equation}  
The orthonormal dual $q$-Krawtchouk functions
satisfy the following recurrence relation, which follows from~\cite[(3.17.3)]{Koekoek}
\begin{equation}
-[-x](1-cq^{x-N})\tilde K_n(\lambda(x))=
- J_{n-1} \tilde K_{n-1}(\lambda(x))+ h_n \tilde K_n(\lambda(x))-J_n \tilde K_{n+1}(\lambda(x)) ,
\end{equation}
where  
\begin{equation}
J_n = -[n-N]\sqrt{\frac{d_{n+1}}{d_n}} ,\qquad h_n = -[n]-cq^{-N}[n-N] .
\label{d-qK-J-h}
\end{equation}
So a similar result holds as in lemma~\ref{lemm1}, except that the single particle eigenvalues are now given by
\[
\epsilon_k=-[-k](1-cq^{k-N})= (q^{-1}+q^{-2}+\cdots+q^{-k}) (1-cq^{k-N}).
\]
Now one can compute the transition amplitude in a spin chain governed by~\eqref{d-qK-J-h}.
As in the previous subsections, the summation part cannot be simplified unless $z^{-[-k](1-cq^{k-N})}=(-1)^k$.
This happens when $q^{-1}=P/Q$ with $P$ and $Q$ positive odd integers, and moreover
$c=-2Q^N$ (or an integer multiple of this), and for the time $t=T=Q^N\pi$. 
Just as in subsection~4.1, one can use (a limit of) the product formula for $q$-Hahn polynomials
in order to rewrite the product of dual $q$-Krawtchouk polynomials as a sum over certain ${}_3\phi_2$-series.
Next, one has to change the order of summation.
For the inner sum over~$k$ there appears, apart from various $q$-shifted factorials, the quotient
$\frac{1-cq^{2k-N}}{1-cq^{-N}}$ (due to the weight function), and this has to be rewritten as
\[
\frac{1-cq^{2k-N}}{1-cq^{-N}} = \frac{(cq^{1-N};q)_{2k}}{(cq^{-N};q)_{2k}} = 
\frac{(c^{1/2}q^{1-N/2},-c^{1/2}q^{1-N/2};q)_{k}}{(c^{1/2}q^{-N/2},-c^{1/2}q^{-N/2};q)_{k}}.
\]
Then, the inner sum reduces to a special case of a very-well-poised ${}_6\phi_5$-series,
which can be summed according to~\cite[(II.20)]{Gasper}.
The final result is:
\begin{align}
f_{r,s}(T)& = \frac{1}{\sqrt{d_r d_s }} \frac{(cq^{1-N},-1;q)_N}{(cq^{1-N};q^2)_N}
\sum_{m=0}^N \frac{(q^{r-N},q^{s-N};q)_m}{(q,-q^{1-N},q^{-N};q)_m} (cq^{1-N};q^2)_m (-c)^{-m} \nn\\
&\times q^{-m(m-1)/2+(N-r-s)m}\, {}_3 \phi_2 \left[ \begin{array}{c} q^{-r},q^{-s},q^{-m} \\ q^{N-r-m+1},q^{N-s-m+1} \end{array}; q,cq^{N+2} \right].
\end{align}
The case $(r,s)=(N,0)$ yields the simplest expression:
\begin{equation}
f_{N,0}(T)= \frac{(-1;q)_N\; (-c)^{N/2} q^{-N(N-1)/4} }{(cq^{1-N};q^2)_N} .
\end{equation}
Again, one can verify that this expression never assumes the value 1 (for $N>1$), 
so perfect state transfer is not possible.

\subsection{$q$-Hahn, dual $q$-Hahn and $q$-Racah polynomials}

The remaining $q$-hypergeometric discrete orthogonal polynomials (with a finite support) are more complicated,
and the notation becomes quite heavy.
We shall give some details regarding the $q$-Racah polynomials, and then some of the results for $q$-Hahn or dual $q$-Hahn polynomials
will follow by taking an appropriate limit.

Let $N$ be fixed and assume $0<q<1$. The $q$-Racah polynomials $R_n(\mu(x);\alpha,\beta,\gamma,\delta;q)$ are polynomials
of degree $n$ in $\mu(x)=q^{-x}+\gamma\delta q^{x+1}$, and involve four parameters $\alpha$, $\beta$, $\gamma$ and $\delta$:
\begin{equation}
R_n(\mu(x)) \equiv R_n(\mu(x);\alpha,\beta,\gamma,\delta;q)=
\,{}_4\phi_3 \left[ \begin{array}{c} q^{-n} ,\alpha\beta q^{n+1}, q^{-x},\gamma\delta q^{x+1} \\ 
\alpha q, \beta\delta q, \gamma q \end{array}; q,q \right], 
\end{equation} 
where $n = 0,1, \ldots ,N$.
One of the following relations should hold: $\alpha q=q^{-N}$, $\beta\delta q=q^{-N}$ or $\gamma q=q^{-N}$.
Here, we shall assume that $\beta\delta q=q^{-N}$, so in the following we put $\delta=\beta^{-1}q^{-N-1}$.

The $q$-Racah polynomials satisfy the orthogonality relation:
\begin{equation}
\sum_{x=0}^N w(x) R_m(\mu(x);\alpha,\beta,\gamma,\delta;q) R_n(\mu(x);\alpha,\beta,\gamma,\delta;q) = d_n \delta_{mn},
\end{equation}
where
\begin{equation}
w(x)= \frac{(\alpha q, \beta\delta q, \gamma q, \gamma\delta q;q)_x}{ (q, \alpha^{-1}\gamma\delta q, \beta^{-1}\gamma q, \delta q;q)_x}
\frac{1-\gamma\delta q^{2x+1}}{1-\gamma\delta q} (\alpha\beta q)^{-x},
\end{equation}
and
\begin{equation}
d_n=\frac{(\alpha\beta q^2, \beta\gamma^{-1};q)_N}{(\alpha\beta\gamma^{-1}q,\beta q;q)_N}
\frac{(q,\alpha\beta\gamma^{-1}q,\alpha\beta q^{N+2},\beta q ;q)_n}{(q^{-N},\alpha q,\alpha\beta q,\gamma q ;q)_n}
\frac{1-\alpha\beta q}{1-\alpha\beta q^{2n+1}} (\beta^{-1}\gamma q^{-N})^n.
\end{equation}
Certain conditions should hold for positivity of the weight function, e.g.\ $0<\alpha<q^{-1}$, $0<\beta<q^{-1}$ and $\gamma>q^{-N}$ (then
automatically $\delta>q^{-N}$);
another option is $\alpha>q^{-N}$, $\beta>q^{-N}$ and $0<\gamma<q^{-1}$.
The orthonormal $q$-Racah functions $\tilde R_n(\mu(x)) = R_n(\mu(x))\sqrt{w(x)/d_n}$ satisfy the
following recurrence relation (see~\cite[(3.2.3)]{Koekoek})
\begin{equation}
-[-x](1-\gamma\delta q^{x+1})\tilde R_n(\mu(x))=
- J_{n-1} \tilde R_{n-1}(\mu(x))+ h_n \tilde R_n(\mu(x))-J_n \tilde R_{n+1}(\mu(x)) ,
\end{equation}
where  
\begin{equation}
J_n = -\frac{A_n}{1-q}\sqrt{\frac{d_{n+1}}{d_n}} ,\qquad h_n = -\frac{A_n+C_n}{1-q} ,
\label{qR-J-h}
\end{equation}
with
\begin{align}
A_n&= \frac{(1-\alpha q^{n+1})(1-\alpha\beta q^{n+1})(1-\beta\delta q^{n+1})(1-\gamma q^{n+1})}{(1-\alpha\beta q^{2n+1})(1-\alpha\beta q^{2n+2})},\nn\\
C_n&= \frac{q(1-q^{n})(1-\beta q^{n})(\gamma-\alpha\beta q^{n})(\delta-\alpha q^{n})}{(1-\alpha\beta q^{2n})(1-\alpha\beta q^{2n+1})}.
\end{align}
We have once again a similar result as in lemma~\ref{lemm1}, but with the single particle eigenvalues given by
\[
\epsilon_k=-[-k](1-\gamma\delta q^{k+1})= (q^{-1}+q^{-2}+\cdots+q^{-k}) (1-\frac{\gamma}{\beta}q^{k-N}).
\]
The next item to work out is the computation of the transition amplitude in a spin chain governed by~\eqref{qR-J-h}.
As in the previous subsections, the summation part cannot be simplified unless $z^{-[-k](1-\frac{\gamma}{\beta}q^{k-N})}=(-1)^k$.
This can occur when $q^{-1}=P/Q$ with $P$ and $Q$ positive odd integers, and moreover
$\gamma=2Q^N\beta$ (or an integer multiple of this), and for the time $t=T=Q^N\pi$. 
The actual computation of $f_{r,s}(T)$ then involves the following steps:
\begin{itemize}
\item 
In the sum
\[
\sum_{k=0}^N w(k) R_s(\mu(k)) R_r(\mu(k)) = \sum_{k=0}^N w(k) R_k(\mu(s)) R_k(\mu(s))
\]
the product of two $q$-Racah polynomials $R_k(\mu(s)) R_k(\mu(r))$ is written as a single sum over ${}_{10}\phi_9$ series, using~\cite[(8.3.1)]{Gasper}.
\item 
Exchanging the order of the two summations, the inner sum can be performed using the very-well-poised ${}_6\phi_5$ summation theorem~\cite[(II.21)]{Gasper}.
This finally leads to a single sum expression of ${}_{10}\phi_9$ series.
\end{itemize}
This last expression reads:
\begin{align}
&f_{r,s}(T) = \frac{(\beta^{-1}\gamma q^{1-N},-1;q)_N}{\sqrt{d_r d_s }(\beta^{-1}\gamma q^{1-N};q^2)_N}
\sum_{m=0}^N \frac{(q^{r-N},q^{s-N},\alpha^{-1}\beta^{-1}q^{-N-r-1},\alpha^{-1}\beta^{-1}q^{-N-s-1} ;q)_m}{(q, \alpha^{-1}\beta^{-1}\gamma q^{-N}, \beta^{-1}q^{-N}, \alpha^{-1}\beta^{-1}q^{-N-1},-q^{1-N},q^{-N};q)_m}  \nn\\
& \times q^m(\beta^{-1}\gamma q^{1-N};q^2)_m \ {}_{10}\phi_9 \Biggl[ \begin{array}{c} \alpha\beta q^{N-m+1}, q \sqrt{\alpha\beta q^{N-m+1}}, 
-q \sqrt{\alpha\beta q^{N-m+1}}, \beta q^{N-m+1}, \\  \sqrt{\alpha\beta q^{N-m+1}}, -\sqrt{\alpha\beta q^{N-m+1}}, \alpha q, \gamma q, \alpha\beta q^{N+2},
\alpha\beta q^{N+r+2-m}, \end{array} \nn\\
& \qquad\qquad\qquad \begin{array}{c} \alpha\beta\gamma^{-1} q^{N-m+1}, q^{-m}, q^{-r}, q^{-s}, \alpha\beta q^{r+1}, \alpha\beta q^{s+1} \\
\alpha\beta q^{N+s+2-m}, q^{N-r+1-m}, q^{N-s+1-m}  \end{array}; q,\beta^{-1}\gamma q^{N+2} \Biggr].
\label{f-racah}
\end{align}
This is quite an impressive expression.
Note that when the sender is at site~$0$ ($s=0$), the ${}_{10}\phi_9$ series collapses, and the remaining sum reduces to a terminating ${}_{4}\phi_3$ series.
In the simple case that $(r,s)=(N,0)$, this simplifies further and yields:
\begin{equation}
f_{N,0}(T)= \frac{(-1;q)_N\; (\gamma/\beta)^{N/2} q^{-N(N-1)/4}}{(\beta^{-1}\gamma q^{1-N};q^2)_N} \sqrt{ 
\frac{(-1)^N (\alpha q, \beta q, \gamma q, \alpha\beta\gamma^{-1}q;q)_N}{(\alpha\beta q^2, \alpha\beta q^{N+1};q)_N} }.
\end{equation}
Unfortunately, one can again verify that this expression never reaches the value 1, 
so perfect state transfer is not possible.

Observe that the $q$-Hahn polynomials $Q_n(q^{-x};\alpha,\beta;q)$ are obtained from the $q$-Racah polynomials in the limit $\gamma\rightarrow 0$.
Without giving all details here, let us mention that in this case the spectrum is of the form~\eqref{eigvals}, and that for 
$t=T=Q^N \pi$ (where $q^{-1}=P/Q$, $P$ and $Q$ positive odd integers), the correlation function is obtained from~\eqref{f-racah}
by taking the same limit. 
In particular, one has in this case:
\begin{equation}
f_{N,0}(T)= (-1;q)_N \sqrt{ \frac{(\alpha q, \beta q;q)_N}{(\alpha\beta q^2, \alpha\beta q^{N+1};q)_N} (\alpha q)^N }.
\end{equation}

In a similar way, the dual $q$-Hahn polynomials $R_n(\mu(x);\gamma,\delta,N;q)$ are obtained from the $q$-Racah polynomials
in the limit $\alpha\rightarrow 0$ (and using $\beta=q^{-N-1}/\gamma$).
The spectrum is of the form
\[
\epsilon_k = -[k](1-\gamma\delta q^{k+1}).
\]
Again under special conditions ($q^{-1}=P/Q$, $P$ and $Q$ positive odd integers; $\gamma\delta=2P^N$; $t=T=Q^N\pi$),
the correlation function is obtained from~\eqref{f-racah} under the same limit.
For the special case $(r,s)=(N,0)$, one finds:
\begin{equation}
f_{N,0}(T)= \frac{(-1;q)_N}{(\gamma\delta q^2;q^2)_N} \sqrt{ (\gamma q, \delta q;q)_N (\gamma q)^N }.
\label{dqHahn}
\end{equation}
A particular simple expression is obtained in this case if both of the parameters $\gamma$ and $\delta$ of the dual
$q$-Hahn polynomials are equal. Then~\eqref{dqHahn} yields
\begin{equation}
f_{N,0}(T)\Big\vert_{\delta=\gamma}= \frac{(-1;q)_N}{(-\gamma q;q)_N} (\gamma q)^{N/2}.
\end{equation}
However, for the allowed range of parameters (for $0<q<1$: $0<\gamma<q^{-1}$ and $0<\delta<q^{-1}$; or $\gamma>q^{-N}$ and $\delta>q^{-N}$)
the expression is always less than~$1$. So once again, perfect state transfer is not achievable.

\section{Summary and conclusions}

In this paper, we have considered linear spin chains with a nearest-neighbour hopping interaction as models for quantum communication.
We have studied the time evolution of single fermion states in such a spin chain.
Certain special spin chains allow perfect state transfer~\cite{Albanese2004,Christandl2004,Kay2009},
and these systems can be related to Krawtchouk polynomials or dual Hahn polynomials.
Following the success of these two initial systems, it was a logical step to study spin chain models related to other
discrete orthogonal systems. This was in fact the topic of an earlier paper~\cite{Chakrabarti2010}.
In that paper, no new models for perfect state transfer were discovered, but the orthogonal polynomial approach
allowed the explicit computation of the correlation function (or transition amplitude) for the known models.
Furthermore, the models were also approached from a group theoretical point of view.

The next logical step is then to study spin chain models related to discrete orthogonal polynomials of $q$-hypergeometric type.
This was exactly the topic of the present paper.
The main feature is that the spin chain data $J_k$ and $h_k$ in~\eqref{Ham2} are now given not by ordinary numbers but by
certain $q$-numbers (of course, since $q$ is a positive parameter, we end up with real numbers again once $q$ is fixed).
We have now investigated all possible spin chain models with an interaction matrix~\eqref{Ham-M} coinciding with
the Jacobi matrix of a discrete orthogonal system of $q$-hypergeometric type.
The fascinating outcome is that one new spin chain model for perfect state transfer has been discovered this way,
in relation to $q$-Krawtchouk polynomials. 
The details of this model and the actual computations have been described in detail in section~3.
A general feature of all the models given in this paper is that the single fermion eigenvalues $\epsilon_j$
are no longer linear in $j$ (as for the Krawtchouk case, see~\cite{Christandl2004}) or quadratic in $j$
(as for the dual Hahn case, see~\cite{Christandl2004}). Instead, the single fermion eigenvalues 
are of the form $\epsilon_j=q^{-1}+q^{-2}+\cdots+q^{-j}$ (or a multiple of this).
As a consequence the correlation function $f_{r,s}(t)$, involving summations over $e^{-it\epsilon_j}$ (see~\eqref{frsz}),
can no longer be simplified in general. 
However, when $q$ is a certain rational number, we have still managed to compute and simplify the 
correlation function for specific values of the time~$t$.
It is in this context that the new model for perfect state transfer was encountered.

For the remaining models based on a Jacobi matrix of affine $q$-Krawtchouk polynomials, quantum $q$-Krawtchouk polynomials, dual $q$-Krawtchouk polynomials,
$q$-Hahn polynomials, dual $q$-Hahn polynomials or $q$-Racah polynomials, perfect state transfer is not achieved.
For all these additional models, we have computed the correlation function (in the cases where simplification takes place).
These computations on their own have some interest, as some remarkable $q$-series manipulations can be performed.
Furthermore, these models might still be worth considering in cases where high-fidelity transfer is of importance.

It is worth mentioning the difference with some recent work on quantum state transfer in a $q$-deformed chain~\cite{Innocente2009}.
In~\cite{Innocente2009}, the algebraic relations for the fermion operators $a_k$, $a^\dagger_k$ in~\eqref{Ham2} themselves are $q$-deformed.
Then, it is rather the representation theory of $q$-deformed algebras that plays a role.
Here, the operators in~\eqref{Ham2} are ordinary undeformed fermion operators. But the fermion chain data, i.e.\
the strengths $J_k$ and $h_k$, are given by $q$-numbers.

\section*{Acknowledgments}
This research was supported by project P6/02 of the Interuniversity Attraction Poles Programme (Belgian State -- 
Belgian Science Policy), and E.I.\ Jafarov acknowledges the support of a Postdoc Fellowship within this Programme.

\end{document}